\documentclass{PoS}

\title{$B_K$ in unquenched QCD using improved staggered fermions}

\ShortTitle{$B_K$ in unquenched QCD}

\author{\speaker{Jongjeong Kim} \\
  Department of Physics and Astronomy,
  Seoul National University, Seoul, 151-747, South Korea \\
  E-mail: \email{rvanguard@phya.snu.ac.kr}}

\author{Taegil Bae \\
  Department of Physics and Astronomy,
  Seoul National University, Seoul, 151-747, South Korea \\
  E-mail: \email{esrevinu@phya.snu.ac.kr}}

\author{Weonjong Lee \\
  Frontier Physics Research Division and
  Center for Theoretical Physics, \\
  Department of Physics and Astronomy, Seoul National University,
  Seoul, 151-747, South Korea \\
  E-mail: \email{wlee@phya.snu.ac.kr}}

\author{Stephen R.~Sharpe \\
  Department of Physics, University of Washington, Seattle,
  WA 98195-1560, USA \\
  E-mail: \email{sharpe@phys.washington.edu}}

\abstract{ We present preliminary results for $B_K$ calculated using
improved staggered fermions with a mixed action (HYP-smeared
staggered valence quarks and AsqTad staggered sea quarks).
We investigate the effect of non-degenerate quarks on $B_K$ and
attempt to estimate the ${\cal O}(a^2)$ effect due to
non-Goldstone pions in loops.
We fit the data to continuum partially quenched chiral perturbation
theory.
We find that the quality of fit for $B_K$ improves if we
include non-degenerate quark mass combinations.
We also observe, however, that the fitting curve deviates from the data
points in the light quark mass region.
This may indicate the need to include taste-breaking in
pion loops.
}

\FullConference{XXIV International Symposium on Lattice Field Theory\\
  July 23-28 2006\\
  Tucson Arizona, US}

\begin{document}

\section{Introduction}
\label{sec:intro}
Indirect CP violation in the neutral kaon system has long been
established experimentally. It is parameterized
by $\epsilon$, which is measured to high precision.
In the standard model, $\epsilon$ is given by
\begin{eqnarray}
  \epsilon &=& C_\epsilon \ \exp(i\pi/4) \ {\rm Im}\lambda_t
  \ X \ \hat{B}_K  \\
  X &=& {\rm Re} \lambda_c [ \eta_1 S_0(x_c) - \eta_3 S_3(x_c,x_t) ]
  - {\rm Re} \lambda_t \eta_2 S_0(x_t) \\
  \lambda_i &=& V_{is}^* V_{id}, \qquad x_i = m_i^2 / M_W^2 \\
  C_\epsilon &=& \frac{G_F^2 F_K^2 m_K M_W^2}{6 \sqrt{2} \pi^2 \Delta M_K}
\end{eqnarray}
where $V_{ij}$ is the CKM matrix element and $S_i$ are the
Inami-Nam functions.
The kaon bag parameter, $B_K$, is defined by
\begin{eqnarray}
  B_K &=& \frac{\langle \bar{K}_0 | [\bar{s} \gamma_\mu (1-\gamma_5) d]
    [\bar{s} \gamma_\mu (1-\gamma_5) d] | K_0 \rangle }{
    \frac{8}{3} \langle \bar{K}_0 | \bar{s} \gamma_\mu\gamma_5 d | 0 \rangle
    \langle 0 | \bar{s} \gamma_\mu\gamma_5 d | K_0 \rangle }
    \\
  \hat{B}_K &=& C(\mu) B_K(\mu)
  \\
  C(\mu) &=& \alpha_s(\mu)^{-\frac{\gamma_0}{2 b_0}}
  [ 1 + \alpha_s(\mu) J_3 ]
\end{eqnarray}
where $\hat{B}_K$ is renormalization-group invariant.
Hence, a precise determination of $B_K$ constrains the CKM matrix---any
violation of the unitarity of this matrix would require new physics
beyond the standard model.
Therefore, there has been a long-standing attempt to calculate
$B_K$ using lattice QCD and other methods.
Staggered fermions have two advantages for a calculation of $B_K$.
First, they preserve an axial $U(1)$ symmetry which constrains the
desired matrix element to have a chiral expansion similar to that
of the continuum, and, second, they are cheap to simulate compared
to formulations with exact (or almost exact) chiral symmetry
(domain wall and overlap fermions). A drawback is the need to
take the fourth-root of the fermion determinant when generating
configurations, leading to unphysical effects for $a\ne 0$, and
the need to use a fitting function containing such effects.
Clearly one must also assume the validity of the rooting trick
in the continuum limit.

Unimproved staggered fermions, however, are not suitable for
a precision calculation. They suffer from
(1) large scaling violations, (2) large taste
symmetry breaking, and (3) large perturbative corrections in matching
factors.
In order to alleviate these problems, there have been a number of
proposals to improve staggered fermions.
Two of them have been used extensively in the lattice community: (1)
AsqTad staggered fermions and (2) HYP/$\overline{\rm Fat7}$ staggered
fermions.
It turns out that HYP staggered fermions reduce the taste symmetry
breaking \cite{ref:wlee:0} as well as one-loop perturbative
corrections \cite{ref:wlee:1} more efficiently
than AsqTad staggered fermions,
and have much smaller scaling violations~\cite{ref:wlee:2}. 
In other words, HYP staggered fermions offer significant
advantages over AsqTad staggered fermions.
Here we perform a numerical study using valence HYP staggered fermions.
We use the unquenched MILC lattices ($N_F=2+1$), in which the
sea quarks are AsqTad staggered fermions, so that we are using
a ``mixed action''. 
The details of the simulation parameters are given in Table
\ref{tab:par:num}.

Although the one-loop contributions to matching factors
with HYP fermions are small,
the unknown two- and higher-loop contributions lead to a significant
uncertainty when one aims at a precision calculation.
We aim to address this problem using both two-loop calculations
and non-perturbative renormalization.

%
%------------------------------------
% parameters for the numerical study
%------------------------------------
\begin{table}[t!]
\begin{center}
\begin{tabular}{ c | c }
\hline 
parameter & value \\ \hline 
$\beta$ & 6.76 (unquenched QCD) \\
sea quarks & 2+1 flavor AsqTad staggered fermions \\ 
valence quarks & HYP staggered fermions \\ 
$1/a$ & 1.588(19) GeV \\ 
geometry & $20^3 \times 64$ \\ 
\# of confs & 640 \\ 
sea quark mass & $a m_u = a m_d = 0.01$, $a m_s = 0.05$ \\ 
valence quark mass & 0.01, 0.015, 0.02, 0.025, 0.03,
0.035, 0.04, 0.045, 0.05 \\ 
$Z_m$ & $\approx 1.0$ \\ \hline
\end{tabular}
\end{center}
\caption{Parameters for the numerical study}
\label{tab:par:num}
\end{table}
\section{$B_K$ for degenerate quarks ($m_x = m_y$)}
\label{sec:bk:deg}
This numerical study is a follow-up to the work
in Ref.~\cite{ref:wlee:4}.
%---------------
% fitting range
%---------------
We place U(1) noise sources at $t=0$ and $t=26$.
These couple to the pseudo-Goldstone pion (spin-taste $\gamma_5 \otimes \xi_5$),
while projecting against all non-Goldstone pions.
When we use this U(1) noise source (at $t=0$) for two-point correlation
functions, we observe a noticeable contamination from
excited states when $0 \le t < 10$.
Hence, we choose the fitting range to be $10 \le t \le 15$ in order to
exclude contamination from both sources.
%

%------------------------------------------
% B_K staggered chiral perturbation theory
%------------------------------------------
In this study, we fit our data to the prediction
of continuum partially quenched chiral perturbation theory.
The result for $N_F=2+1$ flavors was given in Ref. \cite{ref:sharpe:1}.
For degenerate valence quarks ($m_x = m_y$) it takes the form
\begin{eqnarray}
B_K &=& c_1 \bigg( 1 + \frac{1}{48\pi^2 f^2} 
\bigg[ I_{\rm conn} + I_{\rm disc} \bigg] \bigg) 
+ c_2 m^2_{xy} + c_4 m^4_{xy}\,,
\label{eq:bk:deg}
\\
I_{\rm conn} &=& 6 m^2_{xy} \tilde{l}( m^2_{xy} )
- 12 l (m^2_{xy})\,,
\qquad\qquad
I_{\rm disc} =0
\,,
\end{eqnarray}
where $f = 132$ MeV and
\begin{equation}
l(X) = X \log (\frac{X}{\Lambda^2})\,,\qquad
\tilde{l}(X) = - \Big[ \log (\frac{X}{\Lambda^2}) + 1 \Big]\,, 
\end{equation}
%
%\begin{eqnarray}
%l(X) &=& X \log (\frac{X}{\Lambda^2}) + \mbox{Finite Volume Correction} \\
%\tilde{l}(X) &=& - \Big[ \log (\frac{X}{\Lambda^2}) + 1 \Big] + 
%\mbox{Finite Volume Correction}
%\end{eqnarray}
%
up to finite volume corrections.
The notation for meson masses is
\begin{eqnarray}
m^2_U &=& 2\mu m_{u} \ ,
\qquad m^2_S = 2 \mu m_s \ ,
\qquad m^2_\eta = ( m^2_U + 2 m^2_S ) / 3 \\
m^2_X &=& 2\mu m_{x} \ ,
\qquad m^2_Y = 2\mu m_{y} \ ,
\qquad m^2_{xy} = \mu (m_x + m_y)
\end{eqnarray}
where $m_u = m_d \ne m_s$ are sea quark masses and
$m_x$, $m_y$ are valence quark masses.
\begin{figure}[t!]
\begin{center}
\includegraphics[width=0.60\textwidth]{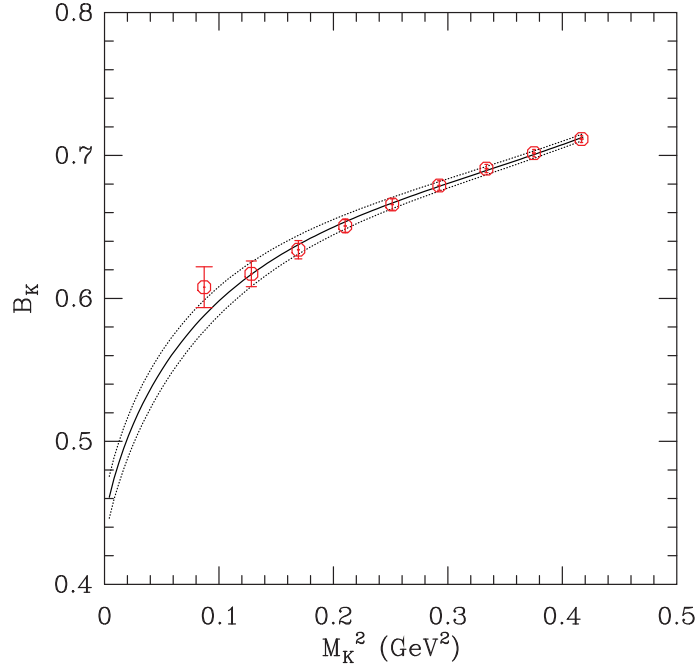}
\end{center}
\caption{$B_K$ versus $m_K^2$ for degenerate quark mass combinations
  ($m_x = m_y$)}
\label{fig:bk:deg}
\end{figure}
In Fig.~\ref{fig:bk:deg}, we plot $B_K$ as a function of $M_K^2$ for
degenerate valence quarks ($m_x = m_y$).
We fit the data to the form in eq.~\ref{eq:bk:deg}, with
the cut-off scale chosen to be $\Lambda = 4\pi f$.
The fitting results are summarized in Table \ref{tab:bk:deg}.
The $\chi^2$ is rather high, considering  that the
data are correlated and we do not include the full covariance matrix.
In particular, we observe that the curve 
does not fit the lightest data point very well.

One possible explanation for the mismatch as small quark mass
is that it is due to
taste-breaking in pion loops. As noted in
Ref.~\cite{ref:sharpe:1}, non-Goldstone pions dominate
the loop contributions which lead to curvature at small $m_K^2$,
and this reduces the expected curvature away from the continuum limit.
This prediction is qualitatively consistent
with what we observe.
However, this behavior could also due to the finiteness of the volume,
which also becomes important only for small $m_K^2$.
Clearly this needs further investigation.

\section{$B_K$ for non-degenerate quarks ($m_x \ne m_y$)}
\label{sec:bk:nd}
A more stringent test of the applicability of chiral perturbation
theory is provided using non-degenerate quarks. The continuum
prediction is~\cite{ref:sharpe:1}
\begin{eqnarray}
B_K &=& c_1 \bigg( 1 +
\frac{1}{48\pi^2 f^2} \Big[ I_{\rm conn} + I_{\rm disc} \Big] 
\bigg)
+ c_2 m_{xy}^2
+ c_3 \frac{ (m_X^2-m_Y^2)^2 }{ m_{xy}^2 }
+ c_4 m_{xy}^4
\nonumber \\ & & \hspace*{+10mm}
\label{eq:bk:nd}
\end{eqnarray}
The contribution from quark-connected diagrams is
\begin{eqnarray}
I_{\rm conn} &=& 6 m_{xy}^2 \tilde{l}(m_{xy}^2)
- 3 l(m_X^2) \Big( 1 + \frac{m_X^2}{m_{xy}^2} \Big)
- 3 l(m_Y^2) \Big( 1 + \frac{m_Y^2}{m_{xy}^2} \Big)
\,,
\end{eqnarray}
while that from diagrams involving a hairpin vertex is
\begin{eqnarray}
I_{\rm disc} &=& ( I_X + I_Y + I_\eta ) / m_{xy}^2
\\
I_X &=& \tilde{l}(m_X^2)
\frac{(m_{xy}^2+m_X^2)(m_U^2-m_X^2)(m_S^2-m_X^2)}{m_\eta^2-m_X^2}
%\nonumber \\ & &
- l(m_X^2)
\frac{(m_{xy}^2+m_X^2)(m_U^2-m_X^2)(m_S^2-m_X^2)}{(m_\eta^2-m_X^2)^2}
\nonumber \\ & &
- l(m_X^2)
\frac{2 (m_{xy}^2+m_X^2)(m_U^2-m_X^2)(m_S^2-m_X^2)}
{(m_Y^2-m_X^2)(m_\eta^2-m_X^2)}
\nonumber \\ & &
- l(m_X^2)
\frac{ (m_U^2-m_X^2)(m_S^2-m_X^2)
-(m_{xy}^2+m_X^2)(m_S^2-m_X^2)
-(m_{xy}^2+m_X^2)(m_U^2-m_X^2)}
{ m_\eta^2-m_X^2 }
\\
I_Y &=& I_X( m_X^2 \leftrightarrow m_Y^2 )
\\
I_\eta &=& l(m_\eta^2)
\frac{ (m_X^2-m_Y^2)^2(m_{xy}^2+m_\eta^2)(m_U^2-m_\eta^2)(m_S^2-m_\eta^2) }
{ (m_X^2-m_\eta^2)^2 (m_Y^2-m_\eta^2)^2 }
\,.
\end{eqnarray}
\begin{table}[t!]
\begin{center}
\begin{tabular}{| c | c | c | c |}
\hline
parameters & unit & average & error \\
\hline
$c_1$ & 1            & 0.4422    & 0.0145 \\
$c_2$ & $GeV^{-2}$   & $-$1.4616 & 0.1611 \\
$c_3$ & $GeV^{-2}$   & $-$       & $-$ \\
$c_4$ & $GeV^{-4}$   & 1.4674    & 0.1913 \\
\hline
$\chi^2$/d.o.f. & 1 & 0.5396    & 0.3 \\
\hline
\end{tabular}
\end{center}
\caption{Fitting results of $B_K$ for degenerate quark mass
combinations ($m_x = m_y$)}
\label{tab:bk:deg}
\end{table}
\begin{figure}[t!]
\begin{center}
\includegraphics[width=0.60\textwidth]{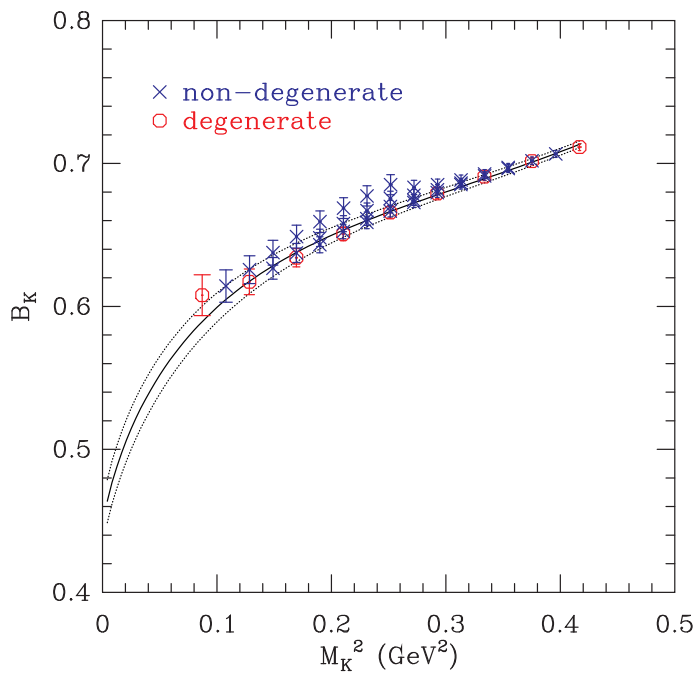}
\end{center}
\caption{$B_K$ versus $m_K^2$ for both degenerate
and non-degenerate quark mass combinations. The fit line
is for the degenerate combinations only.}
\label{fig:bk:nd}
\end{figure}
In Fig.~\ref{fig:bk:nd}, we plot $B_K$ data as a function of $M_K^2$,
including data for both degenerate and non-degenerate valence quarks.
We fit the data to the form in eq.~\ref{eq:bk:nd}.
The results of the fit are summarized in Table \ref{tab:bk:nd}.

When we compare Table \ref{tab:bk:nd} with Table \ref{tab:bk:deg}, we
observe the following: (1) $\chi^2/{\rm d.o.f.}$ is much smaller 
when we include non-degenerate combinations; 
(2) the low energy constants $c_1$, $c_2$ and $c_4$ are consistent within
statistical uncertainty between the two fits; and (3) for our
choice of the cut-off scale, $\Lambda = 4 \pi f$, the low energy
constant $c_3$ is negligibly small.
This comparison tells us that, for this cut-off,
the effect of non-degenerate quark
masses are dominated by the chiral logarithmic terms contained in
$I_{\rm conn}$ and $I_{\rm disc}$, whereas the $c_3$ term plays no role.
Overall, we learn that, while we can in principle
determine $c_1$, $c_2$ and $c_4$ from the degenerate
combinations alone, we have more confidence in the
result when we include non-degenerate combinations.
Furthermore, if we are to include higher order terms in
chiral perturbation theory, as might be required at the heavier
quark masses, the additional non-degenerate points will be 
essential.

The primary reason to study non-degenerate valence quarks is
in order to extrapolate to the physical kaon, in which the
non-degeneracy is almost maximal. The data shows a systematic
increase in $B_K$ with increasing non-degeneracy.
This effect is
statistically significant though numerically small.
When we have a satisfactory description of our data we will
be able to extrapolate to the physical kaon mass.

\section{Staggered chiral perturbation theory}
\label{sec:stag:chipt}
A shortcoming with our present analysis is that it assumes
that Goldstone and non-Goldstone pions are degenerate.
We know, however, that taste symmetry is broken
at ${\cal O}(a^2)$, leading to splittings 
between pions with different tastes.
A systematic way to incorporate such ${\cal O}(a^2)$ effects 
is provided by staggered chiral
perturbation theory \cite{ref:wlee:3,ref:bernard:4}.
In the case of $B_K$, Van de Water and Sharpe have determined
the next-to-leading-order (NLO) prediction of
staggered chiral perturbation theory~\cite{ref:sharpe:1}.
(In fact, a small modification of this result is necessary to
describe our data, since we are using a mixed action.)
Unfortunately, the result contains 21 unknown low energy
constants, even at a single lattice spacing, 
and these must, at present, be determined
from a fit to the lattice data. (Some can be determined in the
future by calculating other matrix elements, or using non-perturbative
renormalization to determine matching factors.)
This is why we have chosen so many mass combinations (45 at present).

The fitting is further constrained if we know 
the masses of the non-Goldstone pions. These enter into
loops, and, as noted above, lead to a modification of
the curvature at small $m_K^2$. Because of this, in a companion
work, we are calculating all pion masses (as well as those
of other hadrons) using HYP-smeared valence staggered
fermions~\cite{ref:wlee:0}. 
A striking result of that study is that the taste-breaking
effects are very small, much smaller than those with AsqTad
improved staggered fermions. Thus it may be that
we can get away with a continuum-like fit except at
the smallest quark masses.

An exception to this comment concerns the contributions
from quark-disconnected diagrams involving hairpin vertices.
These bring in 
taste-singlet pions composed of (AsqTad) sea-quarks, for which
${\cal}(a^2)$ terms are known to be large.
Fortunately, the masses of these pions are
available from the MILC collaboration studies.

As always when doing chiral extrapolations, it is important
to push to as light a quark mass as possible (within the
constraints of finite volume effects and computer time).
In particular, this allows one to more thoroughly test the
applicability of chiral perturbation theory at the order we work.
Because of this, and to provide more data for fitting, we
are presently extending the calculation to a lighter valence
quark mass ($m=0.005$), which will also increase the total
number of mass combinations to 55.

\begin{table}[t!]
\begin{center}
\begin{tabular}{| c | c | c | c |}
\hline
parameters & unit & average & error \\
\hline
$c_1$ & 1                  & 0.4451    & 0.0149 \\
$c_2$ & $GeV^{-2}$         & $-$1.5049 & 0.1672 \\
$c_3$ & $GeV^{-2}$         & 0.0305    & 0.0083 \\
$c_4$ & $GeV^{-4}$         & 1.5366    & 0.2021 \\
\hline
$\chi^2$/d.o.f. & 1 & 0.1888    & 0.0935 \\
\hline
\end{tabular}
\end{center}
\caption{Fitting results of $B_K$ for non-degenerate quark mass
combinations ($m_x \ne m_y$)}
\label{tab:bk:nd}
\end{table}

\section{Acknowledgment}
\label{sec:ack}
W.~Lee acknowledges with gratitude that the research at Seoul National
University is supported by the KOSEF grant (R01-2003-000-10229-0), by
the KOSEF grant of international cooperative research program, by the
BK21 program, and by the US DOE SciDAC-2 program.
S.~Sharpe is supported in part by US DOE grant
DE-FG02-96ER40956 and by the US DOE SciDAC-2 program.

\end{document}